\documentclass[aps,prl,reprint,superscriptaddress,letterpaperm,showpacs]{revtex4-1}
\usepackage[english]{babel}
\usepackage{amsmath}
\usepackage{amsfonts}
\usepackage{amssymb}
\usepackage{graphicx}
\usepackage{bm}
\usepackage{bbm}
\usepackage{empheq}
\usepackage{epstopdf}
\usepackage{hyperref}

\begin{document}

\title{Extreme Harmonic Generation in Electrically Driven Spin Resonance}
\author{J. Stehlik}
\affiliation{Department of Physics, Princeton University, Princeton, NJ 08544, USA}
\author{M. D. Schroer}
\altaffiliation{Present Address: JILA, National Institute of Standards and Technology and the University of Colorado, Boulder, Colorado 80309, USA}
\affiliation{Department of Physics, Princeton University, Princeton, NJ 08544, USA}
\author{M. Z. Maialle}
\affiliation{Faculdade de Ci\^{e}ncias Aplicadas, Universidade Estadual de Campinas, 13484-350 Limeira, SP, Brazil}
\author{M. H. Degani}
\affiliation{Faculdade de Ci\^{e}ncias Aplicadas, Universidade Estadual de Campinas, 13484-350 Limeira, SP, Brazil}
\author{J. R. Petta}
\affiliation{Department of Physics, Princeton University, Princeton, NJ 08544, USA}
\pacs{71.70.Ej, 73.63.Kv, 76.30.-v, 85.35.Gv}

\begin{abstract}
We report the observation of multiple harmonic generation in electric dipole spin resonance in an InAs nanowire double quantum dot. The harmonics display a remarkable detuning dependence: near the interdot charge transition as many as eight harmonics are observed, while at large detunings we only observe the fundamental spin resonance condition. The detuning dependence indicates that the observed harmonics may be due to Landau-Zener transition dynamics at anticrossings in the energy level spectrum.
\end{abstract}

\maketitle

Electron spin resonance and nuclear magnetic resonance are powerful probes of condensed matter systems \cite{Slichter,Abragam}. Electron spin resonance is widely used for quantum control of spin systems, including lithographically defined quantum dots \cite{HansonRev} and nitrogen-vacancy centers in diamond \cite{NVESR}. Nuclear magnetic resonance has resulted in many discoveries in condensed matter physics, such as dynamic nuclear polarization \cite{DNP}, the Overhauser effect \cite{OverhauserEffect}, and double resonance \cite{DoubleRes}. In a typical spin resonance experiment a static magnetic field $B$ Zeeman splits the spin states, while an oscillating perpendicular magnetic field with frequency $f$ drives coherent spin rotations, known as Rabi oscillations.  Electron spin rotations occur when the resonance condition $h f = g \mu_{\rm B} B$ is met. Here $h$ is Planck's constant, $g$ is the electronic $g$-factor, and $\mu_{\rm B}$ is the Bohr magneton. Driving the spin with a detuning large compared to the on-resonance Rabi frequency greatly suppresses the coherent rotations \cite{RabiFormula}.

While electron spin resonance is usually performed on macroscopic ensembles, where the sample size is large enough to directly detect the magnetization, it is difficult to apply directly to nanoscale systems, such as quantum dots \cite{QDotESR}. This has led to the development of single spin resonance, where oscillating magnetic fields are generated via current carrying striplines \cite{StripLineSpinResonance}, or effective ac magnetic fields are generated by electrically driving the electron in a strong spin-orbit material \cite{RashbaEDSR,GolovachLoss,NowackEDSR,PergeEDSR}, a magnetic field gradient \cite{SlantingPRL,SlantingNatPhys}, or a spatially varying hyperfine field \cite{Laird.ESR.PRL}. In quantum dot experiments spin readout is performed using spin-to-charge conversion, where the dc current through the double quantum dot (DQD) is measured in the presence of microwave driving or a local charge sensor is used to detect a spin-dependent tunneling event in the device \cite{PauliBlockade}.

In this Letter we present a detailed study of electric dipole spin resonance (EDSR) in an InAs nanowire DQD \cite{SpinOrbitQubit}. We observe strong harmonic generation in EDSR at driving frequencies satisfying the condition $n h f = g \mu_{\rm B} B$, with $n$ an integer as large as 8 in our system. Several earlier works showed harmonic generation with $n$ = 2 \cite{HalfFreq1,SchroerPRL,NadjPergeSpectroscopy}, but with little discussion of the result or mechanism. While some theories predict harmonic generation in electrically driven quantum dots \cite{RashbaPRB,Nowak,EvenOdd}, the origin of the harmonics remains unclear, as detailed quantitative measurements are absent.

\begin{figure}[t]
\begin{center}
\includegraphics[width=\columnwidth]{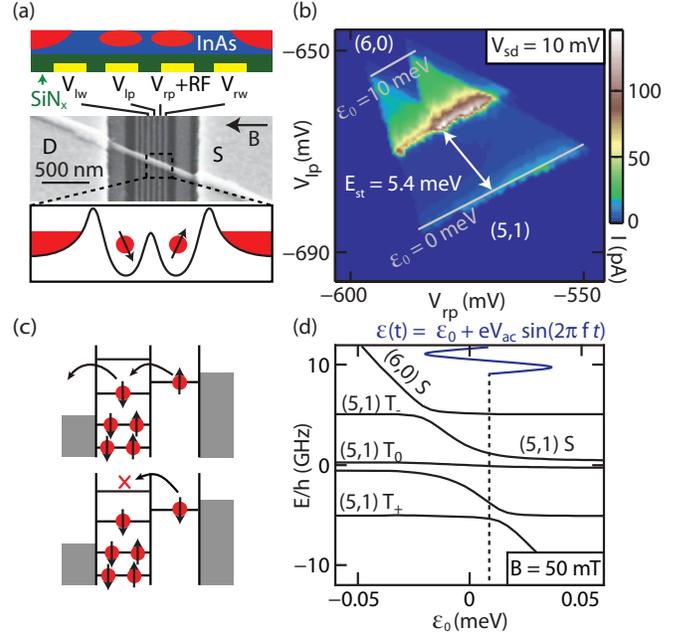}
\caption{(Color online) (a) Scanning electron micrograph illustrating the sample geometry. Negative voltages applied to gates below the nanowire form a DQD.  A magnetic field is applied in the plane of the substrate.  (b) Current $I$ as a function of gate voltages $V_{\rm lp}$ and $V_{\rm rp}$ showing finite bias triangles for the (5,1)$\leftrightarrow$(6,0) charge transition.  Transport is strongly suppressed for $\epsilon_0$ $<$ 5.4 meV due to Pauli blockade. (c) Transport in the sequential tunneling regime. With $\epsilon_0 < E_{\rm st}$, singlet state tunneling is allowed and triplet state tunneling is Pauli blocked. (d) Energy levels as a function of $\epsilon_0$.}
\label{fig1}
\end{center}	
\vspace{-0.6cm}
\end{figure}

\begin{figure*}[t]
\begin{center}
\includegraphics[width=2\columnwidth]{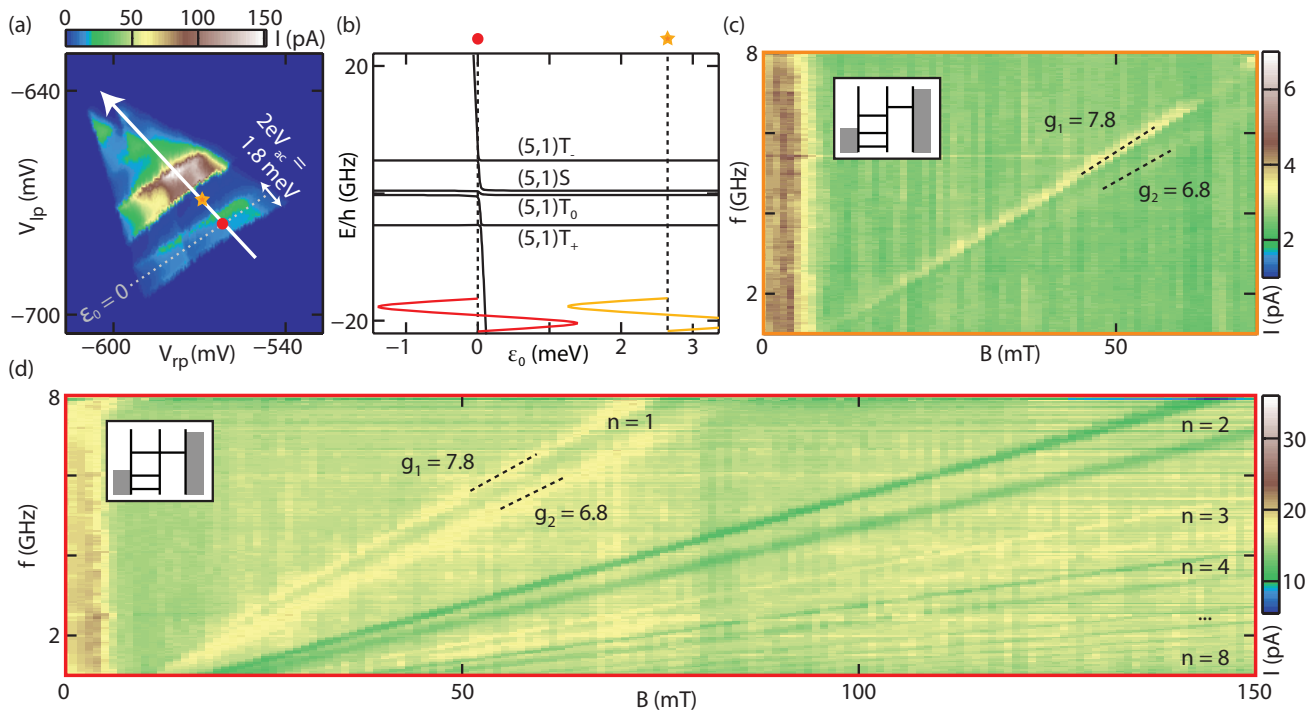}
\caption{(Color online) (a) $I$ as a function of $V_{\rm lp}$ and $V_{\rm rp}$. Microwave driving with $f$ = 4 GHz and $2 eV_{\rm ac}$ = 1.8 meV broadens the features in the charge stability diagram. A white line indicates the detuning axis. (b) Energy level diagram with the ac excitation drawn to scale in the zero detuning region (circle) and the far detuned region (star). (c) $I$ as a function of $B$ and  $f$, taken at $\epsilon_0$ = 2.6 meV, labeled by a star in (a). (d) $I$ as a function of $B$ and $f$, taken near $\epsilon_0$ = 0, indicated by the circle in (a). We observe up to $n = 8$ harmonics in the EDSR response. The approximate level configurations are shown in the insets.}
\label{harm2}
\end{center}
\end{figure*}

A key finding of our measurements is that the presence of harmonics is strongly linked to the detuning of the DQD energy levels. For example, near the interdot charge transition we observe up to $n=8$ harmonics in the EDSR response. In contrast, when far detuned from the interdot charge transition we only observe the fundamental EDSR response with $n$ = 1. Our observations link the presence of harmonics to anti-crossings in the DQD energy level diagram, suggesting the harmonics are caused by Landau-Zener transition dynamics \cite{ShevchenkoReview,BernsNature}. We also observe an even or odd structure in the EDSR response near zero detuning, where odd harmonics result in increased Pauli blockade leakage current, while even harmonics show a suppression of current. Detailed measurements show a strong sinusoidal modulation of the resonant current with detuning. The data provide very clear signatures that can be used for comparison with future theoretical work.

We perform experiments in quantum dots defined along the length of an InAs nanowire [see Fig.\ 1(a)]. A single 50 nm diameter nanowire is placed on top of pre-patterned Ti/Au depletion gates \cite{SchroerNano}. The nanowire and gates are separated by a 20 nm thick layer of $\mathrm{SiN_x}$, which serves as a gate dielectric. We probe spin physics by defining a DQD using the four gates labeled $lw$, $lp$, $rp$, and $rw$.  We then make use of Pauli blockade, where the interdot tunneling process is spin selective.  Figure 1(b) plots the current $I$ as a function of the gate voltages $V_{\rm lp}$ and $V_{\rm rp}$, showing finite-bias-triangles \cite{RevModPhys.75.1} for the (5,1) $\leftrightarrow$ (6,0) electron transition.  Here we label states $(N_{\rm l}, N_{\rm r})$, where $N_{\rm l}$ ($N_{\rm r}$) is the number of electrons in the left (right) dot.  As shown in the top panel of Fig.\ 1(c), with the DQD biased below the singlet-triplet splitting ($E_{\rm st}$), singlet state tunneling is allowed, while current is blocked when the electrons are in a spin-triplet configuration [lower panel of Fig.\ 1(c)] \cite{PauliBlockade,HansonRev}.  The resulting suppression of current due to Pauli blockade is clearly visible in Fig.\ 1(b).  Measurements of a non-zero leakage current in the Pauli blockade regime are indicative of spin dynamics that rotate a Pauli-blocked triplet state to a singlet state. In Fig.\ 1(d) we plot the level diagram of the DQD as a function of detuning $\epsilon_0$ \cite{RevModPhys.75.1}. Interdot tunneling $t_{\rm c}$ couples the singlet states, (5,1)S and (6,0)S, resulting in an anti-crossing. For this device configuration we estimate $t_{\rm c} < k_{\rm B} T_{\rm e}$, where $T_{\rm e}$ = 150 mK is the electron temperature.
The curvature of the (5,1)T$_{\rm 0}$ level is due to the $g$-factor difference between the dots. The (6,0)S state is coupled to the (5,1)T$_{\rm +}$ and T$_{\rm -}$ states by spin-orbit and hyperfine interactions \cite{PettaSeminal,NadjPergeSpectroscopy}.

We set $\epsilon_0$ by changing the voltages $V_{\rm lp}$ and $V_{\rm rp}$ along the detuning axis shown in Fig.\ 2(a) (see the Supplemental Material \cite{SOM}).  EDSR transitions are then generated by applying a microwave drive tone at frequency $f$ to gate $rp$.  As a result, the detuning becomes time-dependent with $\epsilon(t)$ = $\epsilon_0 + eV_{\rm ac}\sin(2 \pi f t)$. This results in broadening of the finite-bias-triangles shown in Fig.\ 2(a), which are measured with $2 eV_{\rm ac}$ = 1.8 meV. EDSR experiments have generally been carried out deep in the Pauli-blocked region, such as at the point labeled by the star in Fig.\ 2(a).  Figure 2(c) shows the leakage current $I$ measured at this point as a function of $f$ and $B$. We observe a peak in $I$ around $B$ = 0, consistent with a hyperfine mediated spin mixing process with root-mean-square nuclear field $B_{\rm N} \sim$ 3.5 mT \cite{Jouravlev,PergeHyperfine}. We also measure an enhancement in $I$ when the electron spin resonance condition $h f = g_{i} \mu_{\rm B} B$ is met, where $g_{i}$ is the $g$-factor on the $i$-th dot. Within the sensitivity limits of our current measurement (noise floor of 100 fA) we do not observe any clear evidence of harmonic generation at large detunings, i.e.\ there is no observable harmonic response for $n$ $>$ 1.

In stark contrast, Fig.\ 2(d) shows $I$ as a function of $B$ and $f$, this time measured at the interdot charge transition where $\epsilon_0$ = 0 [indicated by a circle in Fig.\ 2(a)]. Unlike the data in Fig.\ 2(c), we observe an EDSR response at many harmonics when $n h f = g_{i} \mu_{\rm B} B$, indicating extreme harmonic generation in the DQD device. The EDSR harmonics are particularly strong since the signal weakens only slowly with increasing $n$, demonstrated by the fact that we readily resolve the $n=8$ harmonic. The striking detuning dependence suggests the presence of harmonics in the EDSR response is related to repeated passages through anti-crossings in the energy level diagram near $\epsilon_0 = 0$.

\begin{figure}
\begin{center}
\includegraphics[width=\columnwidth]{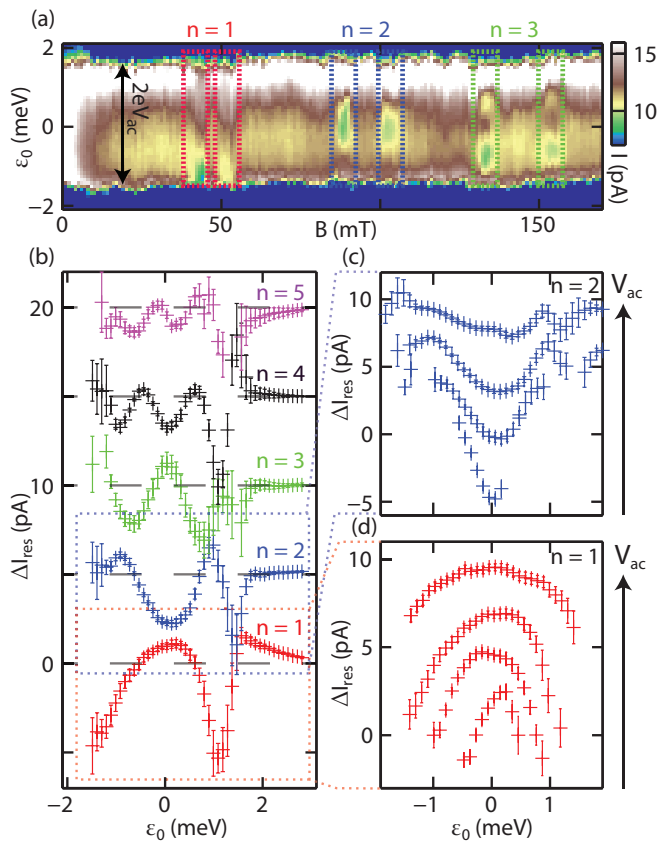}
\caption{(Color online) (a) $I$ as a function of $B$ and $\epsilon_0$ under continuous microwave driving with $f$ = 4.86 GHz and $2 e V_{ac}$ = 2.6 meV. (b) Extracted resonant current $\Delta I_{\rm res}$ for the $n$ = 1 to $n$ = 5 harmonics.  The signal exhibits a sinusoidal-like modulation as a function of $\epsilon_0$.  Successive traces are offset by 5 pA for clarity. $\Delta I_{\rm res}$ for the $n$ = 1 (d) and $n$ = 2 (c) harmonics for $2e V_{ac}$ = 1, 1.8, 2.6, 3.4 meV, respectively. Successive traces are offset by 3 pA for clarity.}
\label{harm3}
\end{center}
\vspace{-0.6cm}
\end{figure}

Additionally, we observe an even or odd structure, wherein the odd harmonics result in an enhancement of the current, meaning spin-blockade is lifted by the action of the microwaves.  The even harmonics, on the other hand, feature a reduction in the current, indicating that spin-blockade is enhanced. Finally, in Fig.\ 2(d) the signal coming from the even harmonics is stronger: while we clearly resolve even harmonics up to $n$ = 8, the odd harmonics become very weak beyond $n$ = 3.

We further study the harmonics by measuring their detuning dependence.  Figure 3(a) shows $I$ as a function of $\epsilon_0$ and $B$ for $f$ = 4.86 GHz and $2 e V_{\rm ac}$ = 2.6 meV.  The resonant signal is clearly visible for the $n$ = 1 through $n$ = 3 harmonics and features clear oscillations as a function of $\epsilon_0$. The off-resonance leakage current is non-zero due to spin-orbit and hyperfine mediated relaxation processes and varies slowly with $\epsilon_0$ \cite{PergeHyperfine}. To examine the resonant response in greater detail we subtract off the non-resonant ``background" to obtain $\Delta I_{\rm res}$, the  change of the leakage current due to resonant microwaves.  We plot this quantity as a function of $\epsilon_0$ for the dot with $g$ = 7.8 in Fig.\ 3(b) (see the Supplemental Material \cite{SOM}). Inside the region where $|\epsilon_0| < e V_{\rm ac}$ the resonant current exhibits sinusoidal oscillations as a function of $\epsilon_0$.  For several values of $\epsilon_0$, $\Delta I_{\rm res}$ is less than zero, meaning the resonant microwaves interfere with other spin-relaxation processes.  With increasing $n$, the frequency of the oscillations in $\Delta I_{\rm res}$ increases, making additional peaks visible.  For odd harmonics, the oscillations feature a local maximum at zero detuning, while for the even harmonics they feature a local minimum.  This behavior is particularly visible as the even or odd structure in Fig.\ 2(d).

In Figs.\ 3(c)--(d) we explore the effect of the amplitude of the applied microwaves on the resonant response by plotting the $n=2$ and $n=1$ resonant signals for $ 2 e V_{\rm ac}$ = 1, 1.8, 2.6 and 3.4 meV.  With increasing driving the amplitude of the oscillations in $\Delta I_{\rm res}$ decreases.  For $n$ = 1 the peak current at zero detuning drops from 2.5 to 0.5 pA.  The frequency of the oscillations also appears to decrease, such that the number of visible peaks for each harmonic remains roughly constant.  Data for $n$ = 3 and $n$ = 4, not shown here, exhibit the same trends.

The measurements presented in Fig.\ 3 provide very clear signatures that can be used for comparison with future theoretical work. A concrete comparison will require a theory that takes into account the noise sources that are inherent to semiconductor DQDs. Of particular interest is detuning noise, which is significant in these devices (typically on the order of 20 $\mu$eV \cite{SchroerParity,KarlFancy}). Additionally, fluctuating nuclear fields and dynamic nuclear polarization processes may be key elements that lead to the interesting even or odd dependence observed in Fig.\ 2(d)\cite{TaylorPRB,koppensModeling}.

To examine the role of the electron occupancy on the harmonics, we repeated the experiments at two other Pauli-blocked interdot transitions. Data from the (5,1) $\leftrightarrow$ (4,2) interdot transition are shown in the Supplemental Material \cite{SOM}. The data sets are consistent with Fig.\ 2: we observe a strong harmonic response near $\epsilon_0 = 0$ with harmonics out to $n=4$ and no harmonic response is observed in the far-detuned regime. Additionally, measurements of $\Delta I_{\rm res}$ as function of $\epsilon_0$ for $f = 4.7 \:\mathrm{GHz}$ and $2 eV_{\rm ac} = 3.2 \:\mathrm{meV}$
show a sinusoidal-like modulation with $\epsilon_0$ and a modulation frequency that increases with $n$ \cite{SOM}.  The data contain the same odd or even structure shown in Fig.\ 2, where odd harmonics result in increased Pauli blockade leakage current, while even harmonics show a suppression of current. Another complete data set acquired at the (7,1) $\leftrightarrow$ (8,0) transition is consistent with these results, indicating that the observed effects are likely independent of the DQD charge occupancy.

In summary, we observe strong harmonic generation in electrically driven spin resonance. Up to eight harmonics are observed near the interdot charge transition. In contrast, at large detunings, only the fundamental EDSR response is observed. The strong detuning dependence illustrated by the data in Fig.\ 2 indicates that Landau-Zener transition physics may play an important role in the generation of the harmonic response observed near $\epsilon_0$ = 0.

We thank E.\ L.\ Rashba and Y.\ Y.\ Liu for helpful discussions. Research at Princeton was supported by the Sloan and Packard Foundations, Army Research Office Grant No.\ W911NF-08-1-0189, DARPA QuEST Grant No.\ HR0011-09-1-0007, and the NSF through DMR-0819860 and DMR-0846341. MZM and MHD acknowledge support from Fapesp and INCT-DISSE/CNPq, Brazil. Research was carried out in part at the Center for Functional Nanomaterials, Brookhaven National Laboratory, which is supported by DOE BES Contract No.\ DE-AC02-98CH10886.


\begin{thebibliography}{37}
\expandafter\ifx\csname natexlab\endcsname\relax\def\natexlab#1{#1}\fi
\expandafter\ifx\csname bibnamefont\endcsname\relax
  \def\bibnamefont#1{#1}\fi
\expandafter\ifx\csname bibfnamefont\endcsname\relax
  \def\bibfnamefont#1{#1}\fi
\expandafter\ifx\csname citenamefont\endcsname\relax
  \def\citenamefont#1{#1}\fi
\expandafter\ifx\csname url\endcsname\relax
  \def\url#1{\texttt{#1}}\fi
\expandafter\ifx\csname urlprefix\endcsname\relax\def\urlprefix{URL }\fi
\providecommand{\bibinfo}[2]{#2}
\providecommand{\eprint}[2][]{\url{#2}}

\bibitem[{\citenamefont{Slichter}(1990)}]{Slichter}
\bibinfo{author}{\bibfnamefont{C.~P.} \bibnamefont{Slichter}},
  \emph{\bibinfo{title}{Principles of magnetic resonance}}
  (\bibinfo{publisher}{Springer-Verlag}, \bibinfo{year}{1990}).

\bibitem[{\citenamefont{Abragam}(1961)}]{Abragam}
\bibinfo{author}{\bibfnamefont{A.}~\bibnamefont{Abragam}},
  \emph{\bibinfo{title}{The principles of nuclear magnetism}}
  (\bibinfo{publisher}{Oxford, Claredon Press}, \bibinfo{year}{1961}).

\bibitem[{\citenamefont{Hanson et~al.}(2007)\citenamefont{Hanson, Kouwenhoven,
  Petta, Tarucha, and Vandersypen}}]{HansonRev}
\bibinfo{author}{\bibfnamefont{R.}~\bibnamefont{Hanson}},
  \bibinfo{author}{\bibfnamefont{L.~P.} \bibnamefont{Kouwenhoven}},
  \bibinfo{author}{\bibfnamefont{J.~R.} \bibnamefont{Petta}},
  \bibinfo{author}{\bibfnamefont{S.}~\bibnamefont{Tarucha}}, \bibnamefont{and}
  \bibinfo{author}{\bibfnamefont{L.~M.~K.} \bibnamefont{Vandersypen}},
  \bibinfo{journal}{Rev. Mod. Phys.} \textbf{\bibinfo{volume}{79}},
  \bibinfo{pages}{1217} (\bibinfo{year}{2007}).

\bibitem[{\citenamefont{Gruber et~al.}(1997)\citenamefont{Gruber, Dräbenstedt,
  Tietz, Fleury, Wrachtrup, and Borczyskowski}}]{NVESR}
\bibinfo{author}{\bibfnamefont{A.}~\bibnamefont{Gruber}},
  \bibinfo{author}{\bibfnamefont{A.}~\bibnamefont{Dräbenstedt}},
  \bibinfo{author}{\bibfnamefont{C.}~\bibnamefont{Tietz}},
  \bibinfo{author}{\bibfnamefont{L.}~\bibnamefont{Fleury}},
  \bibinfo{author}{\bibfnamefont{J.}~\bibnamefont{Wrachtrup}},
  \bibnamefont{and} \bibinfo{author}{\bibfnamefont{C.~v.}
  \bibnamefont{Borczyskowski}}, \bibinfo{journal}{Science}
  \textbf{\bibinfo{volume}{276}}, \bibinfo{pages}{2012} (\bibinfo{year}{1997}).

\bibitem[{\citenamefont{Carver and Slichter}(1953)}]{DNP}
\bibinfo{author}{\bibfnamefont{T.~R.} \bibnamefont{Carver}} \bibnamefont{and}
  \bibinfo{author}{\bibfnamefont{C.~P.} \bibnamefont{Slichter}},
  \bibinfo{journal}{Phys. Rev.} \textbf{\bibinfo{volume}{92}},
  \bibinfo{pages}{212} (\bibinfo{year}{1953}).

\bibitem[{\citenamefont{Overhauser}(1953)}]{OverhauserEffect}
\bibinfo{author}{\bibfnamefont{A.~W.} \bibnamefont{Overhauser}},
  \bibinfo{journal}{Phys. Rev.} \textbf{\bibinfo{volume}{92}},
  \bibinfo{pages}{411} (\bibinfo{year}{1953}).

\bibitem[{\citenamefont{Pound}(1950)}]{DoubleRes}
\bibinfo{author}{\bibfnamefont{R.~V.} \bibnamefont{Pound}},
  \bibinfo{journal}{Phys. Rev.} \textbf{\bibinfo{volume}{79}},
  \bibinfo{pages}{685} (\bibinfo{year}{1950}).

\bibitem[{\citenamefont{Cohen-Tannoudji and Lalo\"e}(1977)}]{RabiFormula}
\bibinfo{author}{\bibfnamefont{C.} \bibnamefont{Cohen-Tannoudji}},
\bibinfo{author}{\bibfnamefont{B.} \bibnamefont{Diu}},
\bibnamefont{and}
  \bibinfo{author}{\bibfnamefont{F.}~\bibnamefont{Lalo\"e}},
  \emph{\bibinfo{title}{Quantum Mechanics Volume One}}
  (\bibinfo{publisher}{Wiley, New York}, \bibinfo{year}{1977}),
  \bibinfo{note}{{Chap.} 4, pp. 386--415}.

\bibitem[{\citenamefont{Morton et~al.}(2005)\citenamefont{Morton, Tyryshkin,
  Ardavan, Porfyrakis, Lyon, and Briggs}}]{QDotESR}
\bibinfo{author}{\bibfnamefont{J.~J.~L.} \bibnamefont{Morton}},
  \bibinfo{author}{\bibfnamefont{A.~M.} \bibnamefont{Tyryshkin}},
  \bibinfo{author}{\bibfnamefont{A.}~\bibnamefont{Ardavan}},
  \bibinfo{author}{\bibfnamefont{K.}~\bibnamefont{Porfyrakis}},
  \bibinfo{author}{\bibfnamefont{S.~A.} \bibnamefont{Lyon}}, \bibnamefont{and}
  \bibinfo{author}{\bibfnamefont{G.~A.~D.} \bibnamefont{Briggs}},
  \bibinfo{journal}{Phys. Rev. Lett.} \textbf{\bibinfo{volume}{95}},
  \bibinfo{pages}{200501} (\bibinfo{year}{2005}).

\bibitem[{\citenamefont{Koppens et~al.}(2006)\citenamefont{Koppens, Buizert,
  Tielrooij, Vink, Nowack, Meunier, Kouwenhoven, and
  Vandersypen}}]{StripLineSpinResonance}
\bibinfo{author}{\bibfnamefont{F.~H.~L.} \bibnamefont{Koppens}},
  \bibinfo{author}{\bibfnamefont{C.}~\bibnamefont{Buizert}},
  \bibinfo{author}{\bibfnamefont{K.~J.} \bibnamefont{Tielrooij}},
  \bibinfo{author}{\bibfnamefont{I.~T.} \bibnamefont{Vink}},
  \bibinfo{author}{\bibfnamefont{K.~C.} \bibnamefont{Nowack}},
  \bibinfo{author}{\bibfnamefont{T.}~\bibnamefont{Meunier}},
  \bibinfo{author}{\bibfnamefont{L.~P.} \bibnamefont{Kouwenhoven}},
  \bibnamefont{and} \bibinfo{author}{\bibfnamefont{L.~M.~K.}
  \bibnamefont{Vandersypen}}, \bibinfo{journal}{Nature (London)}
  \textbf{\bibinfo{volume}{442}}, \bibinfo{pages}{766} (\bibinfo{year}{2006}).

\bibitem[{\citenamefont{Landwehr and Rashba}(1991)}]{RashbaEDSR}
\bibinfo{editor}{\bibfnamefont{G.}~\bibnamefont{Landwehr}} \bibnamefont{and}
  \bibinfo{editor}{\bibfnamefont{E.~I.} \bibnamefont{Rashba}}, eds.,
  \emph{\bibinfo{title}{Landau Level Spectroscopy}}
  (\bibinfo{publisher}{North-Holland, Amsterdam}, \bibinfo{year}{1991}),
  \bibinfo{note}{{Chap.} 4, pp. 133--206}.

\bibitem[{\citenamefont{Golovach et~al.}(2006)\citenamefont{Golovach, Borhani,
  and Loss}}]{GolovachLoss}
\bibinfo{author}{\bibfnamefont{V.~N.} \bibnamefont{Golovach}},
  \bibinfo{author}{\bibfnamefont{M.}~\bibnamefont{Borhani}}, \bibnamefont{and}
  \bibinfo{author}{\bibfnamefont{D.}~\bibnamefont{Loss}},
  \bibinfo{journal}{Phys. Rev. B} \textbf{\bibinfo{volume}{74}},
  \bibinfo{pages}{165319} (\bibinfo{year}{2006}).

\bibitem[{\citenamefont{Nowack et~al.}(2007)\citenamefont{Nowack, Koppens,
  Nazarov, and Vandersypen}}]{NowackEDSR}
\bibinfo{author}{\bibfnamefont{K.~C.} \bibnamefont{Nowack}},
  \bibinfo{author}{\bibfnamefont{F.~H.~L.} \bibnamefont{Koppens}},
  \bibinfo{author}{\bibfnamefont{Y.~V.} \bibnamefont{Nazarov}},
  \bibnamefont{and} \bibinfo{author}{\bibfnamefont{L.~M.~K.}
  \bibnamefont{Vandersypen}}, \bibinfo{journal}{Science}
  \textbf{\bibinfo{volume}{318}}, \bibinfo{pages}{1430} (\bibinfo{year}{2007}).

\bibitem[{\citenamefont{Nadj-Perge
  et~al.}(2010{\natexlab{a}})\citenamefont{Nadj-Perge, Frolov, Bakkers, and
  Kouwenhoven}}]{PergeEDSR}
\bibinfo{author}{\bibfnamefont{S.}~\bibnamefont{Nadj-Perge}},
  \bibinfo{author}{\bibfnamefont{S.~M.} \bibnamefont{Frolov}},
  \bibinfo{author}{\bibfnamefont{E.~P. A.~M.} \bibnamefont{Bakkers}},
  \bibnamefont{and} \bibinfo{author}{\bibfnamefont{L.~P.}
  \bibnamefont{Kouwenhoven}}, \bibinfo{journal}{Nature (London)}
  \textbf{\bibinfo{volume}{468}}, \bibinfo{pages}{1084}
  (\bibinfo{year}{2010}{\natexlab{a}}).

\bibitem[{\citenamefont{Tokura et~al.}(2006)\citenamefont{Tokura, van~der Wiel,
  Obata, and Tarucha}}]{SlantingPRL}
\bibinfo{author}{\bibfnamefont{Y.}~\bibnamefont{Tokura}},
  \bibinfo{author}{\bibfnamefont{W.~G.} \bibnamefont{van~der Wiel}},
  \bibinfo{author}{\bibfnamefont{T.}~\bibnamefont{Obata}}, \bibnamefont{and}
  \bibinfo{author}{\bibfnamefont{S.}~\bibnamefont{Tarucha}},
  \bibinfo{journal}{Phys. Rev. Lett.} \textbf{\bibinfo{volume}{96}},
  \bibinfo{pages}{047202} (\bibinfo{year}{2006}).

\bibitem[{\citenamefont{Pioro-Ladriere
  et~al.}(2008)\citenamefont{Pioro-Ladriere, Obata, Tokura, Shin, Kubo,
  Yoshida, Taniyama, and Tarucha}}]{SlantingNatPhys}
\bibinfo{author}{\bibfnamefont{M.}~\bibnamefont{Pioro-Ladriere}},
  \bibinfo{author}{\bibfnamefont{T.}~\bibnamefont{Obata}},
  \bibinfo{author}{\bibfnamefont{Y.}~\bibnamefont{Tokura}},
  \bibinfo{author}{\bibfnamefont{Y.~S.} \bibnamefont{Shin}},
  \bibinfo{author}{\bibfnamefont{T.}~\bibnamefont{Kubo}},
  \bibinfo{author}{\bibfnamefont{K.}~\bibnamefont{Yoshida}},
  \bibinfo{author}{\bibfnamefont{T.}~\bibnamefont{Taniyama}}, \bibnamefont{and}
  \bibinfo{author}{\bibfnamefont{S.}~\bibnamefont{Tarucha}},
  \bibinfo{journal}{Nat. Phys.} \textbf{\bibinfo{volume}{4}},
  \bibinfo{pages}{776} (\bibinfo{year}{2008}).

\bibitem[{\citenamefont{Laird et~al.}(2007)\citenamefont{Laird, Barthel,
  Rashba, Marcus, Hanson, and Gossard}}]{Laird.ESR.PRL}
\bibinfo{author}{\bibfnamefont{E.~A.} \bibnamefont{Laird}},
  \bibinfo{author}{\bibfnamefont{C.}~\bibnamefont{Barthel}},
  \bibinfo{author}{\bibfnamefont{E.~I.} \bibnamefont{Rashba}},
  \bibinfo{author}{\bibfnamefont{C.~M.} \bibnamefont{Marcus}},
  \bibinfo{author}{\bibfnamefont{M.~P.} \bibnamefont{Hanson}},
  \bibnamefont{and} \bibinfo{author}{\bibfnamefont{A.~C.}
  \bibnamefont{Gossard}}, \bibinfo{journal}{Phys. Rev. Lett.}
  \textbf{\bibinfo{volume}{99}}, \bibinfo{pages}{246601}
  (\bibinfo{year}{2007}).

\bibitem[{\citenamefont{Ono et~al.}(2002)\citenamefont{Ono, Austing, Tokura,
  and Tarucha}}]{PauliBlockade}
\bibinfo{author}{\bibfnamefont{K.}~\bibnamefont{Ono}},
  \bibinfo{author}{\bibfnamefont{D.~G.} \bibnamefont{Austing}},
  \bibinfo{author}{\bibfnamefont{Y.}~\bibnamefont{Tokura}}, \bibnamefont{and}
  \bibinfo{author}{\bibfnamefont{S.}~\bibnamefont{Tarucha}},
  \bibinfo{journal}{Science} \textbf{\bibinfo{volume}{297}},
  \bibinfo{pages}{1313} (\bibinfo{year}{2002}).

\bibitem[{\citenamefont{Flindt et~al.}(2006)\citenamefont{Flindt, S\o{}rensen,
  and Flensberg}}]{SpinOrbitQubit}
\bibinfo{author}{\bibfnamefont{C.}~\bibnamefont{Flindt}},
  \bibinfo{author}{\bibfnamefont{A.~S.} \bibnamefont{S\o{}rensen}},
  \bibnamefont{and}
  \bibinfo{author}{\bibfnamefont{K.}~\bibnamefont{Flensberg}},
  \bibinfo{journal}{Phys. Rev. Lett.} \textbf{\bibinfo{volume}{97}},
  \bibinfo{pages}{240501} (\bibinfo{year}{2006}).

\bibitem[{\citenamefont{Laird et~al.}(2009)\citenamefont{Laird, Barthel,
  Rashba, Marcus, Hanson, and Gossard}}]{HalfFreq1}
\bibinfo{author}{\bibfnamefont{E.~A.} \bibnamefont{Laird}},
  \bibinfo{author}{\bibfnamefont{C.}~\bibnamefont{Barthel}},
  \bibinfo{author}{\bibfnamefont{E.~I.} \bibnamefont{Rashba}},
  \bibinfo{author}{\bibfnamefont{C.~M.} \bibnamefont{Marcus}},
  \bibinfo{author}{\bibfnamefont{M.~P.} \bibnamefont{Hanson}},
  \bibnamefont{and} \bibinfo{author}{\bibfnamefont{A.~C.}
  \bibnamefont{Gossard}}, \bibinfo{journal}{Semicond. Sci. Technol.}
  \textbf{\bibinfo{volume}{24}}, \bibinfo{pages}{064004}
  (\bibinfo{year}{2009}).

\bibitem[{\citenamefont{Schroer et~al.}(2011)\citenamefont{Schroer, Petersson,
  Jung, and Petta}}]{SchroerPRL}
\bibinfo{author}{\bibfnamefont{M.~D.} \bibnamefont{Schroer}},
  \bibinfo{author}{\bibfnamefont{K.~D.} \bibnamefont{Petersson}},
  \bibinfo{author}{\bibfnamefont{M.}~\bibnamefont{Jung}}, \bibnamefont{and}
  \bibinfo{author}{\bibfnamefont{J.~R.} \bibnamefont{Petta}},
  \bibinfo{journal}{Phys. Rev. Lett.} \textbf{\bibinfo{volume}{107}},
  \bibinfo{pages}{176811} (\bibinfo{year}{2011}).

\bibitem[{\citenamefont{Nadj-Perge et~al.}(2012)\citenamefont{Nadj-Perge,
  Pribiag, van~den Berg, Zuo, Plissard, Bakkers, Frolov, and
  Kouwenhoven}}]{NadjPergeSpectroscopy}
\bibinfo{author}{\bibfnamefont{S.}~\bibnamefont{Nadj-Perge}},
  \bibinfo{author}{\bibfnamefont{V.~S.} \bibnamefont{Pribiag}},
  \bibinfo{author}{\bibfnamefont{J.~W.~G.} \bibnamefont{van~den Berg}},
  \bibinfo{author}{\bibfnamefont{K.}~\bibnamefont{Zuo}},
  \bibinfo{author}{\bibfnamefont{S.~R.} \bibnamefont{Plissard}},
  \bibinfo{author}{\bibfnamefont{E.~P. A.~M.} \bibnamefont{Bakkers}},
  \bibinfo{author}{\bibfnamefont{S.~M.} \bibnamefont{Frolov}},
  \bibnamefont{and} \bibinfo{author}{\bibfnamefont{L.~P.}
  \bibnamefont{Kouwenhoven}}, \bibinfo{journal}{Phys. Rev. Lett.}
  \textbf{\bibinfo{volume}{108}}, \bibinfo{pages}{166801}
  (\bibinfo{year}{2012}).

\bibitem[{\citenamefont{Rashba}(2011)}]{RashbaPRB}
\bibinfo{author}{\bibfnamefont{E.~I.} \bibnamefont{Rashba}},
  \bibinfo{journal}{Phys. Rev. B} \textbf{\bibinfo{volume}{84}},
  \bibinfo{pages}{241305} (\bibinfo{year}{2011}).

\bibitem[{\citenamefont{Nowak et~al.}(2012)\citenamefont{Nowak, Szafran, and
  Peeters}}]{Nowak}
\bibinfo{author}{\bibfnamefont{M.~P.} \bibnamefont{Nowak}},
  \bibinfo{author}{\bibfnamefont{B.}~\bibnamefont{Szafran}}, \bibnamefont{and}
  \bibinfo{author}{\bibfnamefont{F.~M.} \bibnamefont{Peeters}},
  \bibinfo{journal}{Phys. Rev. B} \textbf{\bibinfo{volume}{86}},
  \bibinfo{pages}{125428} (\bibinfo{year}{2012}).

\bibitem[{\citenamefont{Osika et~al.}(2013)\citenamefont{Osika, Szafran, and
  Nowak}}]{EvenOdd}
\bibinfo{author}{\bibfnamefont{E.~N.} \bibnamefont{Osika}},
  \bibinfo{author}{\bibfnamefont{B.}~\bibnamefont{Szafran}}, \bibnamefont{and}
  \bibinfo{author}{\bibfnamefont{M.~P.} \bibnamefont{Nowak}},
  \bibinfo{journal}{Phys. Rev. B} \textbf{\bibinfo{volume}{88}},
  \bibinfo{pages}{165302} (\bibinfo{year}{2013}).

\bibitem[{\citenamefont{Shevchenko et~al.}(2010)\citenamefont{Shevchenko,
  Ashhab, and Nori}}]{ShevchenkoReview}
\bibinfo{author}{\bibfnamefont{S.}~\bibnamefont{Shevchenko}},
  \bibinfo{author}{\bibfnamefont{S.}~\bibnamefont{Ashhab}}, \bibnamefont{and}
  \bibinfo{author}{\bibfnamefont{F.}~\bibnamefont{Nori}},
  \bibinfo{journal}{Phys. Rep.} \textbf{\bibinfo{volume}{492}},
  \bibinfo{pages}{1} (\bibinfo{year}{2010}).

\bibitem[{\citenamefont{Berns et~al.}(2008)\citenamefont{Berns, Rudner,
  Valenzuela, Berggren, Oliver, Levitov, and Orlando}}]{BernsNature}
\bibinfo{author}{\bibfnamefont{D.~M.} \bibnamefont{Berns}},
  \bibinfo{author}{\bibfnamefont{M.~S.} \bibnamefont{Rudner}},
  \bibinfo{author}{\bibfnamefont{S.~O.} \bibnamefont{Valenzuela}},
  \bibinfo{author}{\bibfnamefont{K.~K.} \bibnamefont{Berggren}},
  \bibinfo{author}{\bibfnamefont{W.~D.} \bibnamefont{Oliver}},
  \bibinfo{author}{\bibfnamefont{L.~S.} \bibnamefont{Levitov}},
  \bibnamefont{and} \bibinfo{author}{\bibfnamefont{T.~P.}
  \bibnamefont{Orlando}}, \bibinfo{journal}{Nature (London)}
  \textbf{\bibinfo{volume}{455}}, \bibinfo{pages}{51} (\bibinfo{year}{2008}).

\bibitem[{\citenamefont{Schroer and Petta}(2010)}]{SchroerNano}
\bibinfo{author}{\bibfnamefont{M.~D.} \bibnamefont{Schroer}} \bibnamefont{and}
  \bibinfo{author}{\bibfnamefont{J.~R.} \bibnamefont{Petta}},
  \bibinfo{journal}{Nano Lett.} \textbf{\bibinfo{volume}{10}},
  \bibinfo{pages}{1618} (\bibinfo{year}{2010}).

\bibitem[{\citenamefont{van~der Wiel et~al.}(2002)\citenamefont{van~der Wiel,
  De~Franceschi, Elzerman, Fujisawa, Tarucha, and
  Kouwenhoven}}]{RevModPhys.75.1}
\bibinfo{author}{\bibfnamefont{W.~G.} \bibnamefont{van~der Wiel}},
  \bibinfo{author}{\bibfnamefont{S.}~\bibnamefont{De~Franceschi}},
  \bibinfo{author}{\bibfnamefont{J.~M.} \bibnamefont{Elzerman}},
  \bibinfo{author}{\bibfnamefont{T.}~\bibnamefont{Fujisawa}},
  \bibinfo{author}{\bibfnamefont{S.}~\bibnamefont{Tarucha}}, \bibnamefont{and}
  \bibinfo{author}{\bibfnamefont{L.~P.} \bibnamefont{Kouwenhoven}},
  \bibinfo{journal}{Rev. Mod. Phys.} \textbf{\bibinfo{volume}{75}},
  \bibinfo{pages}{1} (\bibinfo{year}{2002}).

\bibitem[{\citenamefont{Petta et~al.}(2005)\citenamefont{Petta, Johnson,
  Taylor, Laird, Yacoby, Lukin, Marcus, Hanson, and Gossard}}]{PettaSeminal}
\bibinfo{author}{\bibfnamefont{J.~R.} \bibnamefont{Petta}},
  \bibinfo{author}{\bibfnamefont{A.~C.} \bibnamefont{Johnson}},
  \bibinfo{author}{\bibfnamefont{J.~M.} \bibnamefont{Taylor}},
  \bibinfo{author}{\bibfnamefont{E.~A.} \bibnamefont{Laird}},
  \bibinfo{author}{\bibfnamefont{A.}~\bibnamefont{Yacoby}},
  \bibinfo{author}{\bibfnamefont{M.~D.} \bibnamefont{Lukin}},
  \bibinfo{author}{\bibfnamefont{C.~M.} \bibnamefont{Marcus}},
  \bibinfo{author}{\bibfnamefont{M.~P.} \bibnamefont{Hanson}},
  \bibnamefont{and} \bibinfo{author}{\bibfnamefont{A.~C.}
  \bibnamefont{Gossard}}, \bibinfo{journal}{Science}
  \textbf{\bibinfo{volume}{309}}, \bibinfo{pages}{2180} (\bibinfo{year}{2005}).

\bibitem[{SOM()}]{SOM}
\bibinfo{note}{See Supplemental Material for the relation between gate voltages and $\epsilon_0$, the determination of the electron occupancy, the method for the extraction of $\Delta
  I_{\rm res}$, and data from other interdot charge transitions.}

\bibitem[{\citenamefont{Jouravlev and Nazarov}(2006)}]{Jouravlev}
\bibinfo{author}{\bibfnamefont{O.~N.} \bibnamefont{Jouravlev}}
  \bibnamefont{and} \bibinfo{author}{\bibfnamefont{Y.~V.}
  \bibnamefont{Nazarov}}, \bibinfo{journal}{Phys. Rev. Lett.}
  \textbf{\bibinfo{volume}{96}}, \bibinfo{pages}{176804}
  (\bibinfo{year}{2006}).

\bibitem[{\citenamefont{Nadj-Perge
  et~al.}(2010{\natexlab{b}})\citenamefont{Nadj-Perge, Frolov, van Tilburg,
  Danon, Nazarov, Algra, Bakkers, and Kouwenhoven}}]{PergeHyperfine}
\bibinfo{author}{\bibfnamefont{S.}~\bibnamefont{Nadj-Perge}},
  \bibinfo{author}{\bibfnamefont{S.~M.} \bibnamefont{Frolov}},
  \bibinfo{author}{\bibfnamefont{J.~W.~W.} \bibnamefont{van Tilburg}},
  \bibinfo{author}{\bibfnamefont{J.}~\bibnamefont{Danon}},
  \bibinfo{author}{\bibfnamefont{Y.~V.} \bibnamefont{Nazarov}},
  \bibinfo{author}{\bibfnamefont{R.}~\bibnamefont{Algra}},
  \bibinfo{author}{\bibfnamefont{E.~P. A.~M.} \bibnamefont{Bakkers}},
  \bibnamefont{and} \bibinfo{author}{\bibfnamefont{L.~P.}
  \bibnamefont{Kouwenhoven}}, \bibinfo{journal}{Phys. Rev. B}
  \textbf{\bibinfo{volume}{81}}, \bibinfo{pages}{201305}
  (\bibinfo{year}{2010}{\natexlab{b}}).


\bibitem[{\citenamefont{Petersson et~al.}(2012)\citenamefont{Petersson, McFaul,
  Schroer, Jung, Taylor, Houck, and Petta}}]{KarlFancy}
\bibinfo{author}{\bibfnamefont{K.~D.} \bibnamefont{Petersson}},
  \bibinfo{author}{\bibfnamefont{L.~W.} \bibnamefont{McFaul}},
  \bibinfo{author}{\bibfnamefont{M.~D.} \bibnamefont{Schroer}},
  \bibinfo{author}{\bibfnamefont{M.}~\bibnamefont{Jung}},
  \bibinfo{author}{\bibfnamefont{J.~M.} \bibnamefont{Taylor}},
  \bibinfo{author}{\bibfnamefont{A.~A.} \bibnamefont{Houck}}, \bibnamefont{and}
  \bibinfo{author}{\bibfnamefont{J.~R.} \bibnamefont{Petta}},
  \bibinfo{journal}{Nature (London)} \textbf{\bibinfo{volume}{490}},
  \bibinfo{pages}{380} (\bibinfo{year}{2012}).

\bibitem[{\citenamefont{Schroer et~al.}(2012)\citenamefont{Schroer, Jung,
  Petersson, and Petta}}]{SchroerParity}
\bibinfo{author}{\bibfnamefont{M.~D.} \bibnamefont{Schroer}},
  \bibinfo{author}{\bibfnamefont{M.}~\bibnamefont{Jung}},
  \bibinfo{author}{\bibfnamefont{K.~D.} \bibnamefont{Petersson}},
  \bibnamefont{and} \bibinfo{author}{\bibfnamefont{J.~R.} \bibnamefont{Petta}},
  \bibinfo{journal}{Phys. Rev. Lett.} \textbf{\bibinfo{volume}{109}},
  \bibinfo{pages}{166804} (\bibinfo{year}{2012}).





\bibitem[{\citenamefont{Taylor et~al.}(2007)\citenamefont{Taylor, Petta,
  Johnson, Yacoby, Marcus, and Lukin}}]{TaylorPRB}
\bibinfo{author}{\bibfnamefont{J.~M.} \bibnamefont{Taylor}},
  \bibinfo{author}{\bibfnamefont{J.~R.} \bibnamefont{Petta}},
  \bibinfo{author}{\bibfnamefont{A.~C.} \bibnamefont{Johnson}},
  \bibinfo{author}{\bibfnamefont{A.}~\bibnamefont{Yacoby}},
  \bibinfo{author}{\bibfnamefont{C.~M.} \bibnamefont{Marcus}},
  \bibnamefont{and} \bibinfo{author}{\bibfnamefont{M.~D.} \bibnamefont{Lukin}},
  \bibinfo{journal}{Phys. Rev. B} \textbf{\bibinfo{volume}{76}},
  \bibinfo{pages}{035315} (\bibinfo{year}{2007}).

\bibitem[{\citenamefont{Koppens et~al.}(2007)\citenamefont{Koppens, Buizert,
  Vink, Nowack, Meunier, Kouwenhoven, and Vandersypen}}]{koppensModeling}
\bibinfo{author}{\bibfnamefont{F.~H.~L.} \bibnamefont{Koppens}},
  \bibinfo{author}{\bibfnamefont{C.}~\bibnamefont{Buizert}},
  \bibinfo{author}{\bibfnamefont{I.~T.} \bibnamefont{Vink}},
  \bibinfo{author}{\bibfnamefont{K.~C.} \bibnamefont{Nowack}},
  \bibinfo{author}{\bibfnamefont{T.}~\bibnamefont{Meunier}},
  \bibinfo{author}{\bibfnamefont{L.~P.} \bibnamefont{Kouwenhoven}},
  \bibnamefont{and} \bibinfo{author}{\bibfnamefont{L.~M.~K.}
  \bibnamefont{Vandersypen}}, \bibinfo{journal}{J. Appl. Phys.}
  \textbf{\bibinfo{volume}{101}}, \bibinfo{eid}{081706} (\bibinfo{year}{2007}).

\end{thebibliography}
\end{document}